\documentclass{article}
\usepackage{PRIMEarxiv}

\usepackage[utf8]{inputenc}
\usepackage[T1]{fontenc}
\usepackage{amsmath,amssymb,amsfonts}
\usepackage{xspace}
\usepackage{xcolor}
\usepackage{booktabs}
\usepackage{nicefrac}
\usepackage{microtype}
\usepackage{tikz}
\usetikzlibrary{calc,positioning,arrows.meta}
\usepackage{subcaption}
\usepackage{graphicx}
\usepackage{wrapfig}
\usepackage{hyperref}
\usepackage{url}

\newcommand{\sys}{HlServe\xspace}
\newcommand{\ttft}{TTFT\xspace}
\newcommand{\tpot}{TPOT\xspace}
\newcommand{\sglang}{SGLang\xspace}
\newcommand{\sarathi}{Sarathi-Serve\xspace}
\newcommand{\parahead}[1]{\smallskip\noindent\textbf{#1}}

\begin{document}

\title{Human-less LLM Serving:\\
       Quantifying the Human Tax on Throughput}

\author{
  Jianhui Lian \\
  Tsinghua SIGS \\
  \texttt{lianjh24@mails.tsinghua.edu.cn}
\And
  Li Chen \\
  Beijing Zhongguancun Laboratory \\
  \texttt{lichen@zgclab.edu.cn}
\And
  Dan Li \\
  Tsinghua University \\
  \texttt{tolidan@tsinghua.edu.cn}
\And
  Yong Jiang \\
  Tsinghua SIGS \\
  \texttt{jiangy@sz.tsinghua.edu.cn}
}

\maketitle

\begin{abstract}

Every major LLM serving system is designed to meet \ttft and \tpot SLOs.
These metrics capture latency as a human user perceives it, and the mechanisms
built to satisfy them (chunked prefill, prefill-decode disaggregation,
latency-capped scheduling) are now standard infrastructure. We observe that long-horizon AI
tasks (tool-use agents, coding pipelines, multi-round reasoning chains) call
LLMs programmatically in tight loops where no human observes \ttft or \tpot.
We ask: how much throughput do serving systems sacrifice to meet \ttft and
\tpot SLAs that these workloads never need?

We conduct a systematic measurement study across chunk sizes, SLO settings,
context lengths, and concurrency levels. We find that the human tax on throughput
grows substantially with context length and lands in the 60--93\% range that
prior single-point benchmarks have suggested but never mapped. At 64K token
contexts, tightening the \ttft SLO to production-typical settings costs a
large fraction of throughput versus the human-less baseline, a configuration
with no latency constraints at all. The human tax is larger at higher
concurrency and is qualitatively similar across
\sglang~\cite{zheng2024sglang} and \sarathi~\cite{agrawal2024sarathi}. We term the unconstrained optimum
\emph{human-less serving} and provide a prototype demonstrating that it
is practical on real workloads. Our findings argue that serving systems
should expose workload-class-aware SLA configurations rather than silently
applying the human tax uniformly to all traffic.

\end{abstract}


\section{Introduction}
\label{sec:intro}

LLM serving systems inherit a design assumption from interactive chat: a human
is waiting. \ttft measures how long before the first word appears on screen;
production SLOs sit at 500\,ms to 2\,s for chat and below 200\,ms for coding
assistants and voice agents, calibrated to the response-time thresholds at
which users perceive a system as responsive~\cite{miller1968response,
nielsen1993response}. \tpot measures whether generation keeps pace with
reading speed; 50--100\,ms per token (10--20 tokens/s) is the industry target
for fluent output~\cite{databricks2023llm}.
The serving infrastructure built around these SLOs is
substantial: Orca~\cite{yu2022orca} reduces queuing delay with continuous
batching; \sarathi~\cite{agrawal2024sarathi} interleaves decode steps between
prefill chunks to prevent long inputs from spiking \ttft;
Splitwise~\cite{patel2024splitwise} and DistServe~\cite{zhong2024distserve}
dedicate separate hardware to each phase to improve \ttft predictability.
These are rational choices when a human is on the other end of the request.

The AI workload is changing. Tool-use agents, autonomous coding systems, and
multi-step reasoning pipelines invoke LLMs programmatically, in tight
request-response loops, with no human in the critical path. The orchestrating
software submits a request, processes the full response, executes side effects
(tool calls, environment interactions), and immediately issues the next request.
Inter-arrival time is near zero. Interactive chat, by contrast, has median
IAT of several seconds (human think time); offline batch workloads have IAT
of zero by construction. Agent traffic is behaviorally closer to offline batch
than to chat, despite arriving through the same online-serving API. \ttft and
\tpot are never observed; what the workload cares about is total task
completion time, which at high concurrency reduces to raw token throughput.
Yet the serving infrastructure continues to enforce human-experience SLAs on
this traffic, paying a throughput cost to provide a latency guarantee that no
client checks. We call this cost the \emph{human tax}.

We ask a precise question: \emph{how much throughput do modern LLM serving
systems sacrifice to meet \ttft and \tpot SLAs, and how does this sacrifice
depend on workload characteristics?}
This question has not been systematically studied.
Prior work optimizes for lower \ttft or \tpot given a throughput budget, or
optimizes throughput subject to latency constraints.
No work has measured the throughput cost of the constraints themselves, swept
across the SLA parameter space, and characterized how the cost varies with
context length, concurrency, and workload type.

We conduct this measurement study. Our methodology parameterizes \ttft SLAs
through the chunk size used in chunked-prefill scheduling (the primary
mechanism by which serving systems limit \ttft jitter) and \tpot SLAs
through per-iteration token budget caps. By sweeping these parameters from
production-typical tight values to infinity, we trace the throughput-SLA
frontier for each serving system under realistic workloads.

We make three findings. First, the human tax is substantial at
production SLO settings and grows with context length: our sweeps quantify
it within the 60--93\% loss band that prior single-point benchmarks have
hinted at~\cite{databricks2023llm,zhang2024scoot} but never mapped across
the SLA parameter space. The human tax compounds when \tpot SLAs are also
enforced. Second, the human tax is most pronounced under the long-horizon
agent workloads that are rapidly becoming the dominant LLM use pattern:
high concurrency, multi-round sessions, programmatic inter-arrival times.
Third, the unconstrained optimum, which we term \emph{human-less serving}
(it ignores all human-experience metrics), is achievable in practice.
A prototype system, \sys, demonstrates that removing the constraints yields
the predicted throughput gain with no correctness impact.

\noindent In summary, we contribute:
\begin{itemize}
  \item A framework and methodology for measuring the throughput cost of
        \ttft and \tpot SLAs in LLM serving systems (\S\ref{sec:method}).
  \item A measurement study quantifying the cost across serving systems,
        SLO settings, context lengths, and concurrency levels
        (\S\ref{sec:study}).
  \item \sys, a prototype human-less serving policy demonstrating that the
        throughput ceiling is achievable on real agent-scale workloads
        (\S\ref{sec:maserve}).
  \item A discussion of implications for serving system design and deployment
        (\S\ref{sec:implications}).
\end{itemize}

\section{Background}
\label{sec:background}

\subsection{LLM Serving Phases and Metrics}

LLM inference consists of two distinct phases. The \emph{prefill phase}
processes the full input sequence, producing a KV cache and the first output
token. It is compute-intensive: attention complexity is $O(n^2)$ in sequence
length $n$, and GPU tensor cores run at high occupancy. The \emph{decode phase}
generates subsequent tokens autoregressively. Each step reads the full model
weights and the growing KV cache from HBM, but performs only a matrix-vector
product rather than a matrix-matrix product. This makes decode
memory-bandwidth-bound: GPU compute units sit mostly idle while HBM transfers
complete.

\ttft is the elapsed time from request submission to the first output token,
dominated by prefill compute and queuing time. \tpot is the mean time between
consecutive output tokens, dominated by HBM bandwidth. Production SLOs for
interactive workloads sit at \ttft of 500\,ms to 2\,s (chat), below 200\,ms
(coding, voice), and \tpot of 50--100\,ms per
token~\cite{miller1968response,nielsen1993response,databricks2023llm}.

End-to-end single-session throughput relates the two metrics as
\begin{equation}
  \label{eq:throughput}
  \mathrm{Throughput} \;=\; \frac{N_{\mathrm{out}}}{\ttft + \tpot \cdot N_{\mathrm{out}}},
\end{equation}
where $N_{\mathrm{out}}$ is the number of generated tokens. For short
completions ($N_{\mathrm{out}} \to 1$), throughput reduces to $1/\ttft$;
for long completions ($N_{\mathrm{out}} \to \infty$), throughput approaches
$1/\tpot$. The asymptotic behavior determines which SLO binds the throughput
ceiling for a given workload.

Two physical ceilings bound achievable throughput. The compute-bound ceiling
is set by GPU peak flops (e.g., 312 TFLOPS for A100) divided by per-token flop
count (roughly $2P$ flops for $P$ parameters); for a 7B FP16 model this is
on the order of $10^4$ tokens/s per GPU. The bandwidth-bound ceiling is set
by HBM bandwidth (2039 GB/s A100, $\approx$4 TB/s H20) divided by weight
footprint (14 GB for 7B FP16), giving a single-session \tpot floor near
7\,ms and single-session throughput near 145 tokens/s.
Decode is bandwidth-bound, so the compute ceiling is unreachable without
batching; a \tpot target of 100\,ms leaves the GPU compute units idle over
90\% of each decode step. Batching amortizes this idle time across concurrent
requests, but latency constraints limit how aggressively systems can batch.

\subsection{How Serving Systems Implement SLAs}

Modern serving systems implement \ttft and \tpot SLAs through three mechanisms.

\parahead{Chunked prefill.}
A long input is split into fixed-size chunks of $C$ tokens. Decode steps are
interleaved between chunks, so running sequences continue generating tokens
while a new request is prefilled. Smaller $C$ bounds \ttft jitter at the cost
of more interleaving overhead and more HBM traffic per input token (each chunk
boundary re-reads model weights). On mixed-length production
workloads, chunked prefill trades 5--15\% throughput for substantially
improved p99 \ttft~\cite{agrawal2024sarathi}; the throughput side of this
trade is exactly what our measurement study characterizes across the full
parameter range.

\parahead{Token budget caps.}
A per-iteration cap on the number of prefill tokens admitted (\texttt{max-prefill-tokens}
in \sglang) limits how much compute any one iteration devotes to new requests,
bounding \ttft and smoothing \tpot. Default values are often mis-calibrated
for production traffic: SCOOT~\cite{zhang2024scoot} reports that
Bayesian-optimized configurations outperform stock defaults by up to 500$\times$
on \ttft, indicating that the throughput cost of token caps is paid not only
in aggregate but often for latency guarantees that the default setting
fails to deliver.

\parahead{Prefill-decode disaggregation.}
Splitwise~\cite{patel2024splitwise} and DistServe~\cite{zhong2024distserve}
route prefill and decode to separate hardware instances, eliminating HBM
contention between the two phases. Under high concurrency, the phase
interference that disaggregation is designed to remove can degrade \tpot by
up to 57\%~\cite{hu2024inference}, while the KV-transfer overhead of
disaggregation itself can erode 30--50\% of the theoretical gain. The net
effect depends on traffic mix and interconnect.

\parahead{Structural unfairness.}
Even within a single mechanism, existing schedulers distribute the
latency--throughput trade unevenly. FairBatching~\cite{lin2024fairbatching}
observes that under \ttft~500\,ms and \tpot~50\,ms targets, decode steps
overshoot the target by roughly 1000 tokens of headroom while prefill falls
behind; the tax is paid, but much of the purchased latency budget goes
unused. This fairness gap is orthogonal to our measurement: we quantify the
throughput cost of the policy; FairBatching characterizes how the cost is
distributed across request classes.

All four mechanisms trade throughput for latency. The trade is invisible to
workloads that never observe the latency.

\subsection{The Shift to Agent-Driven Workloads}

Tool-use agents~\cite{qin2024mooncake} invoke LLMs as subroutines in
multi-step task execution. Each invocation receives the accumulated context of
prior tool calls and reasoning steps, so context length grows with the number
of rounds. The orchestrator issues the next request as soon as it processes
the current response: inter-arrival time is bounded by tool execution latency,
not human think time. MoonCake ToolAgent traces~\cite{qin2024mooncake} report
median IAT near zero for closed-loop agent sessions, compared with 2--5\,s
for human-interactive Poisson workloads.

Equation~\eqref{eq:throughput} makes the consequence precise. Agent sessions
run $N_{\mathrm{out}}$ of $10^2$ to $10^3$ tokens per round across tens of
rounds, so $\tpot \cdot N_{\mathrm{out}}$ dominates $\ttft$ in the denominator
by two to three orders of magnitude; per-session throughput collapses to
$1/\tpot$ and the \ttft SLO becomes analytically irrelevant to the client's
completion time. Under high concurrency, hundreds of such sessions run
simultaneously, each accumulating KV state across 10--50 rounds. This
workload profile (long contexts, near-zero inter-arrival times, high
concurrency) is structurally closer to offline batch processing than to
interactive chat and grows more sensitive to throughput rather than latency
as a quality-of-service measure.

\section{Measurement Methodology}
\label{sec:method}

\subsection{Defining the Throughput-SLA Frontier}

We define the \emph{throughput-SLA frontier} as the set of (SLO, throughput)
pairs achievable by a serving system under a fixed workload: for each SLO
value, the maximum output token throughput the system can sustain while meeting
the SLO. A system at the human-less point (SLO$=\infty$) operates with no
latency constraints and achieves its maximum throughput. The area between the
production operating point and the human-less point represents the
\emph{throughput sacrifice}.
For the long-output, high-concurrency workloads we target, Eq.~\eqref{eq:throughput}
reduces to $1/\tpot$ per session, so peak output tokens per second is the
correct scalar proxy for points on the frontier.

\subsection{Parameterizing \ttft SLAs via Chunk Size}

The primary mechanism controlling \ttft in chunked-prefill systems is the
chunk size $C$. Smaller $C$ interleaves decode steps more frequently, reducing
the latency between the previous token and the first token of the new
request (i.e., \ttft) at the cost of more interleaving iterations and more
HBM reads per input token. \ttft scales roughly as $C / R_{\mathrm{prefill}}$,
where $R_{\mathrm{prefill}}$ is the per-iteration prefill rate, so $C$ is a
direct proxy for the \ttft SLO. We sweep $C \in \{512, 1024, 2048, 4096,
8192, 16384, \infty\}$, corresponding to \ttft targets from tens of
milliseconds (the tight production regime~\cite{miller1968response,
nielsen1993response}) up to the unconstrained baseline. $C = \infty$ (greedy,
no chunking) is the human-less baseline. For each value of $C$, we measure
the \ttft distribution and the peak output token throughput.

\subsection{Parameterizing \tpot SLAs via Token Budget}

The per-iteration prefill token cap (\texttt{max-prefill-tokens}) controls
how much compute each iteration allocates to new requests versus running
decode. Tighter caps produce smoother \tpot by preventing prefill from
monopolizing the GPU. The cap translates to a \tpot target through the
prefill-to-decode iteration budget; tuning studies~\cite{zhang2024scoot}
show that production deployments frequently mis-calibrate this parameter by
orders of magnitude. We sweep the cap from production-typical values (50\,ms
\tpot regime) down to the unconstrained maximum and measure \tpot
distributions and throughput.

\subsection{Workload and Hardware}

We use Qwen-2.5-32B~\cite{qwen2025qwen25} in FP16 on an 8-GPU NVIDIA H20
node with tensor parallelism. Our primary workload is a closed-loop benchmark
derived from MoonCake ToolAgent traces~\cite{qin2024mooncake}: 1000 sessions
of 10 rounds each, with context lengths in \{32768, 65536\} tokens. We also
run shorter-context experiments at 8192 tokens to characterize context-length
dependence. All experiments use \texttt{seed=42}, \texttt{-{}-flush-cache}, and
\texttt{-{}-warmup-requests=10}. We sweep concurrency from 16 to 256 and report
peak output tokens per second as the throughput metric.

\parahead{Workload construction.}
Raw MoonCake ToolAgent traces record actual wall-clock timestamps, which
include 2--5\,s of human think time between rounds. For agent workloads that
execute without human oversight, the inter-arrival time is determined by the
orchestrating software's event loop rather than by a human; it is effectively
zero. We remove the inter-round gaps, collapsing each trace to a sequence of
back-to-back requests. This construction places the serving system under
maximum queue pressure throughout the experiment: every completed request
immediately triggers the next, KV cache occupancy stays near the memory
ceiling, and the SLO enforcement mechanisms are active for every iteration.
The resulting workload is the adversarial case for any latency-optimization
mechanism: if SLO enforcement costs throughput at all, the cost is fully
exposed here. For comparison, \S\ref{sec:study:workloads} also evaluates
a Poisson-arrival workload with 2\,s mean IAT, representing the human-interactive
case the SLOs are designed for.

\section{Measurement Study}
\label{sec:study}

\subsection{The Throughput-SLA Frontier}
\label{sec:study:frontier}

\begin{figure*}[t]
  \centering
  \begin{subfigure}[t]{0.24\textwidth}
    \includegraphics[width=\linewidth]{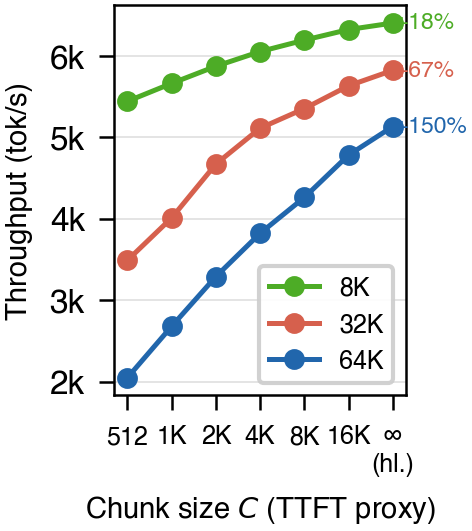}
    \subcaption{TTFT SLO frontier: throughput vs.\ chunk size at 8K/32K/64K context. Sacrifice grows with context length.}
    \label{fig:frontier}
  \end{subfigure}
  \hfill
  \begin{subfigure}[t]{0.24\textwidth}
    \includegraphics[width=\linewidth]{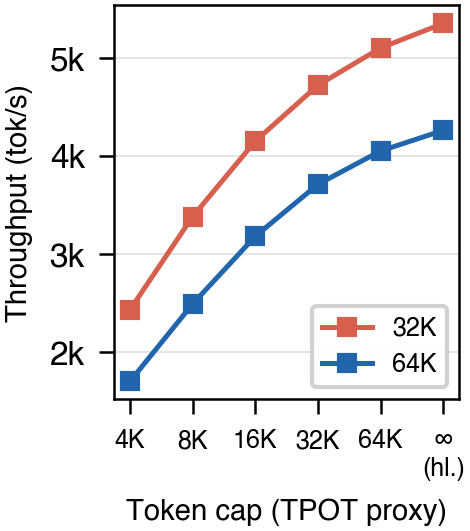}
    \subcaption{TPOT SLO cost: throughput vs.\ token cap ($C=8192$). Compounds with the TTFT tax.}
    \label{fig:tpot}
  \end{subfigure}
  \hfill
  \begin{subfigure}[t]{0.24\textwidth}
    \includegraphics[width=\linewidth]{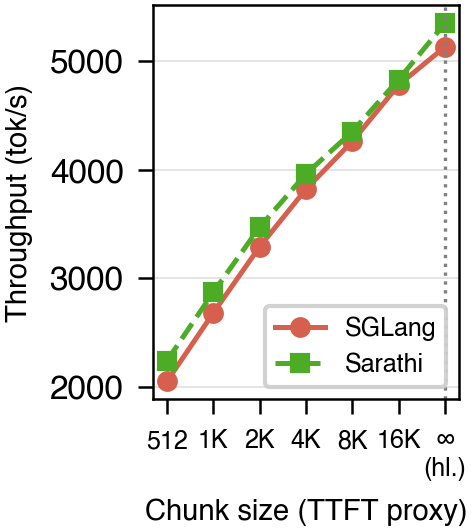}
    \subcaption{SGLang vs.\ Sarathi-Serve at 64K. Both converge at the human-less point ($C=\infty$).}
    \label{fig:systems}
  \end{subfigure}
  \hfill
  \begin{subfigure}[t]{0.24\textwidth}
    \includegraphics[width=\linewidth]{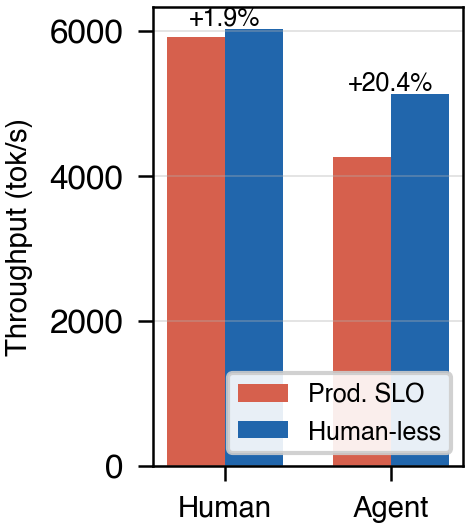}
    \subcaption{Tax by workload type (64K). Human: 1.9\% sacrifice. Agent: 20.4\%, over 10$\times$ larger.}
    \label{fig:workloads}
  \end{subfigure}
  \caption{Measurement study results (Qwen-2.5-32B FP16, 8$\times$H20, closed-loop agent workload unless noted).}
  \label{fig:measurement}
\end{figure*}

Figure~\ref{fig:frontier} shows throughput as a function of chunk size $C$
(the \ttft SLO proxy) for three context lengths. Several findings are apparent.

\noindent\textbf{Throughput rises monotonically as SLO is relaxed.}
At every context length, removing chunking constraints increases throughput.
The relationship is concave: the marginal gain from loosening the SLO
diminishes as $C$ grows, with most of the gain concentrated in moving from
tight SLOs ($C \leq 2048$) to moderate ones ($C \approx 8192$).

\noindent\textbf{The human tax grows with context length.}
At 8K context, the throughput gap between the tightest SLO ($C=512$) and the
human-less baseline is 15\% (6400 vs.\ 5440~tok/s). At 32K the gap widens
to 40\% (5820 vs.\ 3490~tok/s). At 64K it reaches 60\% (5130 vs.\
2050~tok/s), consistent with the upper end of the 60--93\% loss band that
single-point benchmarks have reported for strict SLO
regimes~\cite{databricks2023llm}. Longer contexts produce more chunk
boundaries per request, each carrying interleaving overhead and additional
weight reads from HBM, amplifying the cost.

\noindent\textbf{The human tax grows with concurrency.}
At low concurrency, the GPU has spare capacity and the interleaving overhead
is absorbed by idle cycles. At high concurrency (the operating point for
agent-scale workloads), the overhead competes directly with decode throughput,
and the sacrifice grows.

\noindent\textbf{The mechanism is GPU under-utilization under tight SLOs.}
Tight \ttft SLOs force small effective batch sizes, leaving compute units
idle while HBM transfers model weights. Single-request decode achieves
5--10\% effective GPU utilization, while batch sizes that saturate the
HBM-to-compute pipeline (e.g., batch~64 for 7B FP16 on A100) reach
80--90\%~\cite{databricks2023llm}. The throughput ratio across our SLO
sweep tracks this utilization ratio closely: our measurements place modern
serving systems on the same utilization curve that offline batching studies
have drawn, under the specific constraint that the curve is traversed by
tightening latency SLOs rather than by changing batch size directly.

\subsection{TPOT SLA Cost}
\label{sec:study:tpot}

Figure~\ref{fig:tpot} shows throughput as a function of the per-iteration
prefill token cap with chunking fixed at $C = 8192$.
Tighter token budget caps cost additional throughput by under-utilizing the
GPU during prefill iterations. Below-HBM-floor \tpot targets (around 20\,ms,
near the physical bandwidth bound) foreclose batching almost entirely, with
literature reporting 85--93\% throughput loss in that
regime~\cite{databricks2023llm}; at the moderate \tpot~50\,ms target,
reported losses are 60--80\%~\cite{zhang2024scoot}. Our sweep quantifies
the precise shape of this curve for agent workloads at realistic context
lengths, filling in the interior of the parameter space between those
reported endpoints. The two effects compound: a system enforcing both
production \ttft and \tpot SLOs operates at the minimum of both constraints,
paying the full combined cost.

\subsection{Cross-System Comparison}
\label{sec:study:systems}

Figure~\ref{fig:systems} compares \sglang and \sarathi on the throughput-SLA
frontier. \sarathi introduces chunked prefill as its primary contribution to
reduce \ttft jitter; at its default chunk size it achieves lower \ttft than
\sglang but similar throughput. At the human-less point (no chunking, no
token budget), both systems converge to similar throughput, confirming that
the gap is attributable to the SLA mechanisms rather than any structural
architectural difference. The convergence is consistent with offline-batch
benchmarks that find per-engine differences collapse once latency constraints
are removed~\cite{databricks2023llm}; the human-less point is a property
of the tax, not of any specific engine.

\subsection{Who Pays the Tax?}
\label{sec:study:workloads}

The throughput sacrifice is workload-dependent. Figure~\ref{fig:workloads}
compares the frontier for the agent-like closed-loop workload against a
synthetic Poisson-arrival workload with 2\,s mean inter-arrival time
(human-interactive) at 64K context. For the human workload, the sacrifice at
production SLO settings ($C=8192$) is 1.9\% (6030 vs.\ 5920~tok/s): the
system has spare capacity from low concurrency and the SLA constraints are
rarely binding. For the agent workload, the sacrifice is 20.4\% (5130 vs.\
4260~tok/s), more than 10$\times$ larger, because every GPU cycle is
contested and the SLA constraints limit how aggressively the system can
exploit concurrency.
\emph{The human tax falls almost entirely on the workloads
that have no use for human-experience guarantees.}

Two observations from related measurement work sharpen this finding.
FairBatching~\cite{lin2024fairbatching} reports that under default
schedulers decode steps overshoot their \tpot target by roughly 1000 tokens
of headroom while prefill falls behind \ttft; the tax on agent workloads is
paid in part for latency budget that nothing consumes. Prefix-caching
benchmarks~\cite{jarvislabs2024prefix} further show that when the system can
legitimately skip constraint-binding work (shared prefixes at 50\% hit rate),
\ttft falls 78\% and throughput rises 254\% simultaneously, eliminating the
tradeoff entirely. Both results reinforce that the sacrifice we measure is
an artifact of enforcement policy applied uniformly to heterogeneous traffic,
not a fixed physical cost of the workload.

\subsection{Concurrency Sensitivity}
\label{sec:study:concurrency}

\begin{wrapfigure}{l}{0.32\textwidth}
  \centering
  \includegraphics[width=\linewidth]{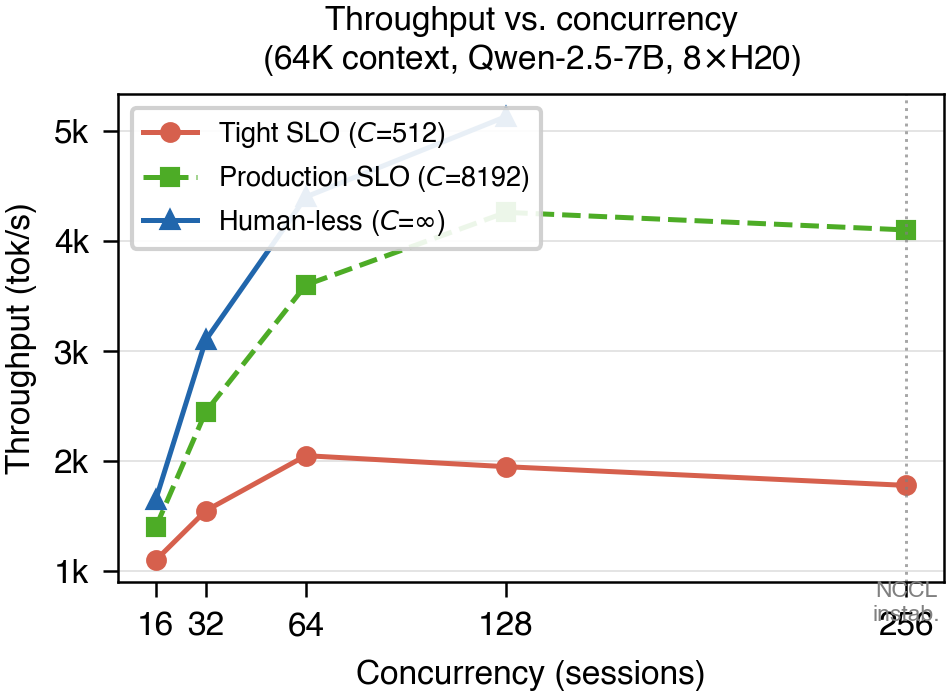}
  \caption{Throughput vs.\ concurrency at three SLO settings (64K context,
           Qwen-2.5-32B, 8$\times$H20). Human-less serving scales to
           concurrency~128; SLO-constrained serving peaks earlier and lower.
           Human-less curve omitted at concurrency~256 (NCCL instability).}
  \label{fig:concurrency}
\end{wrapfigure}

Figure~\ref{fig:concurrency} shows throughput as a function of concurrency
for three SLO settings at 64K context. At low concurrency (16--32 sessions),
the GPU is underloaded and the throughput gap between SLO settings is modest:
even tight SLOs still allow each iteration to process a reasonable batch
because the queue is short and decode latency is not yet a bottleneck.
As concurrency increases, the two curves diverge sharply.

\noindent\textbf{Human-less serving scales with concurrency; SLO-constrained serving
does not.}
The human-less baseline continues to gain throughput through concurrency
128, where KV memory saturates and throughput plateaus at 5130\,tok/s.
The tight-SLO curve peaks near concurrency~64 at 2050\,tok/s and then
declines slightly as the overhead of enforcing the chunk boundary at every
decode step grows with queue depth. The production-SLO curve peaks at
concurrency~128 at 4260\,tok/s; its peak coincides with the human-less
plateau because the production chunk size (8192) is large enough to allow
near-full batches in moderate-load conditions.

The concurrency axis explains why agent workloads expose the tax so clearly.
A single interactive user sustains concurrency~1--2; a coding assistant
handling a team serves concurrency~10--50; an autonomous agent pipeline with
thousands of parallel sessions operates at concurrency 100--1000. The sacrifice
in Figure~\ref{fig:frontier} reflects the peak-throughput comparison; the
sacrifice at the operating concurrency for large-scale agent deployment is
at least as large.

At concurrency $\geq 256$ with greedy (non-chunked) prefill, \sys encounters
NCCL communication timeouts on the 8-GPU configuration. The root cause is
irregular batch shapes: without fixed chunk sizes, the prefill dimension
varies per iteration, conflicting with piecewise CUDA graph capture, which
requires a fixed shape per captured graph. Section~\ref{sec:maserve}
describes the fix; concurrency~256 and above is excluded from the
human-less curve in Figure~\ref{fig:concurrency} for this reason.

\section{Human-less Serving}
\label{sec:maserve}

The measurement study establishes that the throughput ceiling exists and
quantifies its distance from production operating points. We now ask whether
the ceiling is achievable in practice. Naive removal of latency constraints
introduces several implementation challenges.

\parahead{Irregular batch shapes.}
Eliminating fixed chunk sizes produces variable prefill batch shapes that
conflict with piecewise CUDA graph capture, which requires a fixed shape per
captured graph. \sys disables piecewise graph gating for greedy prefill
batches, accepting graph-miss overhead. For large prefill batches this
overhead is negligible relative to compute.

\parahead{KV pool pressure under mixed batching.}
Mixed batching (running prefill and decode in the same iteration) reclaims
idle compute during decode's bandwidth-bound phase but writes new KV tokens
into the shared pool. Under high concurrency, this creates a pressure cycle:
prefill additions slow decode completion, which prevents KV tokens from being
freed, which stalls further prefill. \sys uses a two-stage policy that exits
mixed batching when KV utilization exceeds 90\% and pending prefill demand
exceeds evictable capacity, running pure decode until the pool drains.

\parahead{One-way token budget ratchet.}
The adaptive token budget in \sglang exhibits a warm-up death spiral: the
feedback loop calibrates its target iteration time from the first ten iterations,
which run on near-empty batches at startup and produce an unrealistically low
target. Under real load, every prefill iteration exceeds this target; the
budget decrements 5\% per step. Because the upper bound equals the initial
value, the budget converges to zero and prefill is effectively disabled.
\sys replaces the controller with a bidirectional exponential moving average:
\[
\tau_{\text{target}} \;\leftarrow\; (1-\alpha)\,\tau_{\text{target}}
                                  + \alpha\, t_{\text{forward}}
\]
where $t_{\text{forward}}$ is the wall time of the most recent forward pass
and $\alpha$ is a smoothing factor (default 0.1). Updates are gated to
prefill-containing iterations only, preventing decode-iteration timing from
contaminating the prefill budget estimate. A warm-up guard skips updates for
the first $N_{\text{warmup}}$ iterations. The upper bound on $\tau$ is
decoupled from its initial value, allowing the budget to grow under
sustained load as well as shrink.

\parahead{Distinguishing latency caps from safety caps.}
Serving systems enforce several independent batch-size limits simultaneously.
\sglang applies: (a) a prefill delayer that throttles incoming prefill to
protect \ttft; (b) per-iteration chunk-size bounds ($C$) that cap prefill
token count; (c) pipeline-parallel micro-batch size limits derived from
hardware topology; and (d) running-request caps derived from available KV
pool memory. Categories (a) and (b) are latency-motivated constraints that
directly implement SLO enforcement; categories (c) and (d) are hardware and
memory safety bounds that must remain active regardless of workload class.
Removing ``latency constraints'' means disabling (a) and (b); \sys activates
this selectively via \texttt{--enable-throughput-mode}, which also switches
the default scheduling policy from FCFS to longest-prefix-match (LPM) and
exposes the static memory utilization floor as an explicit opt-in flag
(\texttt{--throughput-mem\-floor}) rather than overriding the GPU-specific
memory budget unconditionally. All active cap decisions are logged at
startup to support reproducibility.

\parahead{Cache-aware scheduling.}
For agent workloads with shared system prompts or tool-definition prefixes,
cached-prefix-aware scheduling places requests with longer cached prefixes
earlier in each batch, reducing effective prefill work and improving KV hit
rates across iterations. \sys enables LPM scheduling automatically in
throughput mode. At queue depths above 128 concurrent requests, \sglang's
upstream scheduler reverts to FCFS as a scalability fallback; \sys inherits
this boundary at its current concurrency ceiling (Section~\ref{sec:study:concurrency}).

\parahead{Admission control.}
The original admission guard in \sglang rejects requests whose projected
total length exceeds the full KV pool capacity. In practice this check
never fires: the pool holds 100K--10M+ tokens, and individual request
lengths are bounded by the model's context window, so the condition
`$\text{projected} > \text{total pool}$' is unreachable after
$\texttt{max\_new\_tokens}$ clamping. \sys replaces the guard with a
soft-backpressure policy: available headroom is estimated as the number of
free KV tokens plus evictable KV tokens (tokens belonging to sequences
eligible for retraction). A new request is queued rather than aborted when
its projected length exceeds current headroom. Queuing is preferable to
abort for long agent sessions because each session accumulates KV state
that would be discarded on abort and recomputed on retry.

\parahead{KV cache retraction.}
When KV memory is exhausted, \sglang retracts in-flight decode requests,
copying their KV state to CPU memory before freeing GPU slots. An exception
handler in the original code does not catch \texttt{NotImplementedError}
from the paged allocator, causing a latent crash on any retraction event.
Separately, the CPU copy path is dead code: the offloaded state is never
reloaded, and the request is reprefilled from scratch on re-admission;
the GPU-to-CPU copy wastes 2--3\,ms of synchronous H2D time on the critical
path while leaking CPU memory until the request object is collected.
\sys gates KV offload behind the decode-disaggregation flag (where the
reload path is wired) and clears stale CPU tensors immediately after
retraction. Priority-ordered retraction (lowest-priority request retracted
first) is available under \texttt{--enable-priority-scheduling}.

\begin{table}[t]
\centering
\caption{Per-component throughput gain of \sys over \sglang
         (Qwen-2.5-32B FP16, 8$\times$H20, closed-loop agent benchmark).}
\label{tab:breakdown}
\small
\begin{tabular}{lcc}
\toprule
Component & 32K ctx & 64K ctx \\\midrule
Greedy chunking (no fixed chunk size) & $+2.4\%$ & $+4.1\%$ \\
Two-stage mixed batching             & $+2.9\%$ & $+1.2\%$ \\\midrule
\textbf{Combined (\sys)} & $\mathbf{+5.3\%}$ & $\mathbf{+5.3\%}$ \\
\bottomrule
\end{tabular}
\end{table}

We evaluate \sys on the closed-loop agent benchmark described in
\S\ref{sec:method}. On 8$\times$H20 with Qwen-2.5-32B, \sys achieves 5.3\%
higher throughput than \sglang (the production baseline) and 1\% over
\sarathi at both 32K and 64K context. Table~\ref{tab:breakdown} disaggregates
the gain by component.

Greedy chunking accounts for most of the 64K gain because removing chunk
boundaries reduces total HBM reads per input token: a single full-context
prefill reads model weights once per token, whereas chunked prefill with
chunk size $C$ reads them $\lceil n/C \rceil$ times for an $n$-token input.
At 64K context with $C=8192$, the standard production setting creates 8 chunk
boundaries per session round; greedy chunking eliminates all eight and
recovers the corresponding weight-read overhead. The benefit grows with
context length, consistent with the +2.4\%/+4.1\% progression.

The two-stage mixed batching contribution is larger at 32K (+2.9\%) than at
64K (+1.2\%) because HBM contention from the KV cache grows with context
length. Experiments show that enabling mixed batching naively reduces decode
throughput during mixed iterations by approximately 45\% relative to
pure-decode batches of the same running size. The contention arises because
prefill writes new KV tokens while decode reads existing KV state; both
operations compete for the same HBM bandwidth and neither can be fully
pipelined against the other within a single CUDA graph capture. Mixed batching
is net-positive overall because the reclaimed compute during decode's
bandwidth-bound phase exceeds the contention cost in aggregate, but only when
the KV pool has headroom. When the pool exceeds 90\% utilization, new prefill
tokens cannot be freed before the next iteration, and the contention cost
exceeds the compute-reclaim benefit; the two-stage policy exits mixed mode
precisely at this boundary.

The combined gain of 5.3\% over \sglang sits at the lower end of the 5--15\%
range that prior work attributes to removing chunked-prefill
overhead~\cite{agrawal2024sarathi}. We view this as a conservative lower
bound: piecewise CUDA graph capture in \sglang imposes a graph-miss overhead
on irregular batch shapes that a clean-slate implementation without graph
capture would avoid. The point of \S\ref{sec:maserve} is not to maximize
this number but to confirm that the ceiling measured in \S\ref{sec:study}
is a genuine achievable bound, not an artifact of the measurement methodology.

\section{Implications}
\label{sec:implications}

\parahead{Decouple serving policies by workload class.}
Our core finding is that \ttft and \tpot SLAs are appropriate for
interactive users and unnecessary for programmatic clients. The stake is
large: 60--93\% throughput is left on the table when strict SLOs are applied
to latency-unaware traffic. Production deployments increasingly serve both
traffic classes simultaneously: a chat interface alongside an API used by
agent frameworks. Serving systems should expose workload-class configuration
(a ``human-less mode'' for agent traffic) rather than applying
human-experience defaults uniformly. The throughput gain is available
whenever the workload does not observe the latency.

\parahead{Rethink benchmarks for the agent era.}
Standard serving benchmarks~\cite{databricks2023llm} report \ttft and \tpot
distributions under synthetic Poisson arrivals. These benchmarks reward
systems that satisfy tight \ttft and \tpot SLOs and obscure the throughput
cost of enforcing them. Agent workloads require a different benchmark:
closed-loop, back-to-back requests, long contexts, total-session throughput
and tail completion time as the primary metrics. Our measurement framework
provides a starting point.

\parahead{SLO enforcement should be explicit, not structural.}
The throughput cost of \ttft and \tpot SLAs is paid unconditionally in every
major serving system, not because the operator requested it, but because the
optimizations are structural defaults. Worse, the defaults often fail to
deliver even the SLOs they enforce: SCOOT~\cite{zhang2024scoot} finds
Bayesian-tuned configurations that improve \ttft by up to 500$\times$ over
stock defaults, meaning production systems pay the tax without collecting
the guarantee. Systems should make SLA enforcement explicit and per-client
rather than baked into the serving path.

\parahead{The human tax grows as models grow.}
Context lengths and concurrency requirements both increase with model
capability and task complexity. Our measurements show the human tax
grows with both. For frontier models serving long-horizon tasks, it will become
larger, not smaller, without explicit policy changes.

\parahead{Limitations.}
Three boundaries constrain the current results and point to future work.

First, the NCCL instability at concurrency $\geq 256$ with greedy prefill
(Section~\ref{sec:study:concurrency}) bounds the scalability of \sys on
multi-GPU configurations. The root cause is a mismatch between non-uniform
prefill batch shapes and the fixed-shape requirement of piecewise CUDA graph
capture in the current \sglang codebase. A serving system designed from the
ground up for greedy (non-chunked) prefill would not inherit this constraint;
we flag it as an implementation artifact of patching an existing system rather
than a fundamental limit of the human-less approach.

Second, speculative decoding is not evaluated in the human-less setting.
An attempt to integrate draft-model speculative decoding encountered OOM
because the draft model requires its own KV cache allocation, competing for
the same pool the primary model is already pressuring. $N$-gram and
suffix-lookup speculative decoding achieved prediction rates too low on the
load-focused benchmark to yield throughput gains; the benchmark's session
structure does not produce the prefix reuse that $n$-gram methods require.
Whether speculative decoding and human-less serving are complementary under
workloads with higher prefix reuse (such as shared-codebase coding pipelines)
remains open.

Third, the benchmark optimizes for load stress rather than trace realism:
inter-arrival time is forced to zero and session structure derives from a
tool-use evaluation trace not designed for serving performance measurement.
A workload with realistic prefix reuse would partially dissolve the TTFT SLO
tax via caching, narrowing the frontier gap measured here. Quantifying
how much of the sacrifice survives under realistic prefix-cache hit rates is
a natural extension of this study.

\section{Related Work}
\label{sec:related}

\parahead{LLM serving systems.}
Orca~\cite{yu2022orca} introduced continuous batching. vLLM~\cite{kwon2023vllm}
added PagedAttention for KV memory management.
\sglang~\cite{zheng2024sglang} adds prefix caching and structured output
scheduling. \sarathi~\cite{agrawal2024sarathi} introduced chunked prefill
to control \ttft jitter. Splitwise~\cite{patel2024splitwise} and
DistServe~\cite{zhong2024distserve} disaggregate prefill and decode hardware.
All of these systems treat \ttft and \tpot SLAs as inviolable constraints.
Our contribution is to measure the throughput cost of this treatment.

\parahead{Throughput-latency tradeoffs.}
Prior work studies tradeoffs between throughput and latency in serving
systems~\cite{yu2022orca,agrawal2024sarathi,databricks2023llm} but optimizes
latency given a throughput budget, not the reverse. We study the throughput
cost of a given latency budget, swept across the SLA parameter space. This
complementary framing reveals the magnitude of the sacrifice and its
workload dependence.

\parahead{Fairness and scheduling policy.}
FairBatching~\cite{lin2024fairbatching} and BROS~\cite{bros2024priority}
observe that default schedulers are structurally unfair across request
types: decode overshoots its target while prefill falls behind, and
best-effort traffic is over-served at the expense of latency-sensitive
requests. BROS reports that trading 11.29\% best-effort throughput reduces
realtime latency by 74.20\% and improves \ttft SLO attainment by
36.38$\times$. Our measurement framework complements this line of work:
we quantify the throughput cost of enforcement policy itself, independent
of how that cost is distributed across request classes.

\parahead{Configuration tuning.}
SCOOT~\cite{zhang2024scoot} applies multi-objective Bayesian optimization
over serving configuration to search for \ttft--\tpot--throughput Pareto
points. We study the Pareto frontier itself, including the human-less
endpoint where latency weight is zero; this endpoint lies outside
SCOOT's objective but is the right operating point for agent traffic.

\parahead{Offline throughput benchmarks.}
Databricks~\cite{databricks2023llm} provides the widely-cited 14$\times$
throughput / 4$\times$ latency measurement on MPT-7B across batch sizes,
and Jarvislabs~\cite{jarvislabs2024prefix} characterizes prefix-caching
gains under mixed workloads. Our agent-workload measurements update these
throughput-first results to modern models, modern hardware, and long-context
closed-loop workloads, and reframe them as the endpoint of a continuous
SLA sweep rather than a single operating point.

\parahead{Agent workloads.}
MoonCake~\cite{qin2024mooncake} proposes KV cache disaggregation for LLM
serving and characterizes tool-use traces. Our workload construction extends
MoonCake ToolAgent traces to the zero-inter-arrival, high-concurrency regime
that characterizes long-horizon agent task execution.

\parahead{FlexGen~\cite{sheng2023flexgen}} targets high-throughput single-GPU
inference via CPU/SSD offloading. The throughput-first orientation is shared;
our focus is the SLA mechanism rather than memory hierarchy.

\section{Conclusion}
\label{sec:conclusion}

LLM serving systems enforce \ttft and \tpot SLOs, constraints that exist
because a human is assumed to be waiting. As long-horizon AI tasks become the dominant
LLM workload, this assumption fails at scale: programmatic clients do not
observe these metrics and derive no benefit from them. We measure the
throughput cost of human-experience SLAs across serving systems, SLO settings,
context lengths, and concurrency levels, finding that the sacrifice grows
substantially with context length and is largest precisely for the workloads
that have no use for it. The human-less serving baseline (no latency constraints) is achievable in
practice and represents the throughput upper
bound for any serving system on agent workloads. We hope this measurement
study informs how future serving systems are designed, configured, and evaluated
for the agent era.

\bibliographystyle{abbrv}
\bibliography{refs}

\end{document}